# Molecular dynamics simulations of interaction between a super edge dislocation and interstitial dislocation loops in irradiated L1$_2$-Ni$_3$Al

Cheng Chen[1,2*], Dongyang Qin[1,2], Yiding Wang[1], Fei Xu[1,2*] and Jun Song[3*]

1. *School of Aeronautics, Northwestern Polytechnical University, Xi'an, Shaanxi 710072, China*
2. *National Key Laboratory of Strength and Structural Integrity, Xi'an 710072, China*
3. *Department of Mining and Materials Engineering, McGill University, Montréal, Québec H3A 0C5, Canada*

**Abstract**

The study employed MD simulations to investigate the interactions between a ⟨110⟩ super-edge dislocation, consisting of the four Shockley partials, and interstitial dislocation loops (IDLs) in irradiated L1$_2$-Ni$_3$Al. Accounting for symmetry breakage in the L1$_2$ lattice, the superlattice planar faults with four distinct fault vectors have been considered for different IDL configurations. The detailed dislocation reactions and structural evolution events were identified as the four partials interacted with various IDL configurations. The slipping characteristics of Shockley partials within the IDLs and the resultant shearing and looping mechanisms were also clarified, revealing distinct energetic transition states determined by the fault vectors after the Shockley partials sweeping the IDL. Furthermore, significant variations in critical resolved shear stress (CRSS) required for the super-edge dislocation to move past the IDL were observed, attributed to various sizes and faulted vectors of enclosed superlattice planar faults in the IDLs. The current study extends the existing dislocation-IDL interaction theory from pristine FCC to L1$_2$ lattice, advances the understanding of irradiation hardening effects in L1$_2$-Ni$_3$Al, and suggests potential applicability to other L1$_2$ systems.

[*] Corresponding authors: E-mail: cheng.chen@nwpu.edu.cn (Cheng Chen);
E-mail: xufei@nwpu.edu.cn (Fei Xu);
E-mail: jun.song2@mcgill.ca (Jun Song), Tel: +1 (514) 398-4592.

# 1. Introduction

The interstitial dislocation loops (IDLs) enclosing planar faults are commonly observed in irradiated FCC metals and alloys [1, 2]. It has been well documented that the interplay between IDLs and moving dislocations may act as an important mechanism for irradiation induced working hardening and plastic instability, significantly influencing the microstructural evolution and mechanical performance of irradiated materials [3-7]. As such, it is of great implication to the embrittlement and failure of structural components used in next generation nuclear reactors, posing challenges to their longevity and performance and consequently attracted great research attention investigating into the subject of IDLs interacting with gliding dislocations [8-11]. Numerous experimental studies have been performed, providing valuable information about the unfaulting processes of IDLs when interacting with dislocations, based on which potential dislocation reaction mechanisms have been proposed [12-17]. Further to these experimental studies, there have also been significant efforts in the past few decades employing atomistic simulations, particularly molecular dynamics (MD) simulations [10, 18-25], given their ability to access the spatial and time resolution not available in typical experiments, in order to reveal fundamental small-scale details of interactions between IDLs and moving dislocations. The combined experimental and atomistic simulation efforts have identified a diverse set of interaction scenarios for the interactions between IDLs and dislocations, including the absorption of IDLs by dislocations, resulting in the formation of superjogs or helical turns, the dislocation-induced unfaulting of IDL into glissile configuration, and the shearing of IDLs [2, 10, 22, 23], among others.

Those prior studies provide critical knowledge for an in-depth understanding of the subject. However, most of the previous research focused on the pristine metal systems, while little attention was directed towards intermetallic compounds possessing ordered superlattice structures. These intermetallic compounds are often essential strengthening agents in metals and alloys. A notable category of such compounds are the FCC-derived $L1_2$ intermetallics, widely existing in superalloys for advanced aerospace applications and chemically complex alloys that are considered as candidate materials for advanced reactor applications [26-31]. For instance, the $L1_2$ - $Ni_3Al$ present in Ni-based superalloys, is coherently embedded in the FCC host metal matrix and primarily responsible for the exceptional high-temperature mechanical properties and corrosion resistance [32-35]. The $L1_2$-strengthened high-entropy alloys (HEAs) were demonstrated to exhibit much improved irradiation resistance, evidenced by the reduced bubble sizes [36, 37] when subjected to irradiation.

Consequent from the little attention devoted, there have been very few studies on the interaction between IDLs and dislocations in $L1_2$-$Ni_3Al$, and more notably, clear lack of systematic nanoscale simulations of dislocation activities therein. Despite the pristine FCC structure being well studied, the knowledge is not directly applicable to the $L1_2$ structure because of the clear distinction in the structural characteristics of IDLs and dislocations in FCC and L1$_2$. In the pristine FCC lattice, there exists only one type of IDL with the extrinsic stacking fault enclosed, whereas in the L1$_2$ lattice, a diverse range of IDL variants exist, attributed to multiple complex planar fault structures, including superlattice intrinsic stacking faults (SISF), anti-phase boundaries (APB), superlattice extrinsic stacking faults (SESF), and complex extrinsic stacking faults (CESF) [2, 16, 38, 39]. Meanwhile, the gliding dislocations in the FCC lattice undergo a splitting process, resulting in the formation of two partial dislocations that are connected by a stacking fault ribbon, while, in contrast, the L1$_2$ structure features super-dislocations that dissociate into two super-partials accompanied by the formation of an APB, subsequently decomposing into two partial dislocations connected either by a CSF or SISF [35]. These additional complexities in the L1$_2$ structure are expected to give rise to new and different mechanisms of sessile IDLs and gliding dislocations, thus necessitating systematic examination.

This study investigates the nanoscale interactions between the super-edge dislocation, and IDLs in a representative L1$_2$-Ni$_3$Al system, employing molecular dynamics (MD) simulations. Accounting for symmetry breakage in the L1$_2$ lattice, multiple orientation combinations of fault vectors of superlattice planar faults within IDLs have been considered to interact with the four Shockley partials that constitutes the super-edge dislocation. The shearing and looping mechanisms, along with the slipping characteristics of four Shockley partials have been uncovered, during which it elucidates the transformation of planar faults enclosed within IDLs and the associated dislocation reactions with diversified transition state structures of IDLs identified. Moreover, the size effect of IDL-super edge interaction has been examined and the corresponding critical stresses required for the super-edge dislocation moving past the IDLs are determined.

## 2. Computational Methodology

The atomistic models constructed are illustrated in **Fig. 1(a-b)**. The model encloses a IDL and a super-edge dislocation residing in the same plane. The supercell has dimensions of $L_{\vec{x}_1} = 603\text{Å}$, $L_{\vec{x}_2} = 490\text{Å}$ and $L_{\vec{x}_3} = 187\text{Å}$, with corresponding orientations being $\vec{x}_1 = [\bar{1}\bar{1}0]$, $\vec{x}_2 = [1\bar{1}2]$, and $\vec{x}_3 = [1\bar{1}1]$ respectively. The supercell is periodic along $\vec{x}_1$ and $\vec{x}_2$, while free along $\vec{x}_3$ direction. In L1$_2$-Ni$_3$Al, the super-edge dislocation has a Burgers vector $[\bar{1}\bar{1}0]$, twice of that of an edge dislocation in FCC-Ni. The super-edge dislocation can be represented by its four constituting Shockley partials of Burgers vectors $\frac{1}{6}\langle 1\bar{1}2 \rangle$, denoted as $B\alpha^1$, $\alpha C^2$, $B\alpha^3$ and $\alpha C^4$ connected by an APB and CSFs, according to the following dislocation dissociation reaction [35, 40, 41]:

$$2BC \rightarrow B\delta^1 + \text{CSF} + \delta C^2 + \text{APB} + B\delta^3 + \text{CSF} + \delta C^4 \quad (1)$$

The superscripts label the indices of the four partials to differentiate them from other equivalent partials. In addition, the positive line sense is represented by the vector $\vec{x}_2 = [1\bar{1}2]$, as indicated by the arrows in **Fig.1**. The IDLs were constructed using a procedure detailed in our previous work [2, 10] and also briefly described below. In the IDL construction, one extra layer of atoms were enclosed in a Frank loop with a Burgers vector of $\frac{1}{3}[1\bar{1}1]$ on the $(1\bar{1}1)$ plane, and then atoms were subjected to a displacement field $U$ produced by the interstitial dislocation loop and the four Shockley partials. Specifically, if we label the atoms in original simulation cell as $C_0$, and the atoms in the inserted layer as $C_1$, the displacement field $U$ for $C_0$ and $C_1$ atoms with coordinates of $\vec{x} = (x_1, x_2, x_3)$ can be given as the follows.

$$U_j(\vec{x}; C_0) = \sum_{l=0}^{l=4} \sum_{i=1}^{i=n_l} -\frac{b_j^l \Omega_i^l(x_1, x_2, x_3)}{4\pi} \tag{2}$$

$$U_j(\vec{x}; C_1) = \sum_{l=1}^{l=4} \sum_{i=1}^{i=n_l} -\frac{b_j^l \Omega_i^l(x_1, x_2, x_3)}{4\pi} + \Delta S_1 \tag{3}$$

The subscripts $j$ =1-3 correspond to the components of $\vec{x}_1$, $\vec{x}_2$ and $\vec{x}_3$ directions, respectively. The Burgers vectors of the inserted interstitial loop is denoted as $\boldsymbol{b^0}$, and the four Shockley partials are denoted as $\boldsymbol{b^l}$ with $l$ =1-4, while $\Omega_i^l$ is the solid angle associated with the $i$-th segment of the $l$-th dislocation. It is important to highlight that the inserted extra atom layer is exempt from their own displacement field and should be sheared by a planar fault vector of $\Delta S_1$ between two neighboring atomic layers to generate the CESF with two-layer CSFs or SESF with two-layer SISFs.

The fault enclosed in the IDL varies depending on $\Delta S_1$, which can assume six possible faulted vectors, i.e., $\boldsymbol{\delta A}, \boldsymbol{\delta B}, \boldsymbol{\delta C}, \boldsymbol{2A\delta}, \boldsymbol{2B\delta}$ and $\boldsymbol{2C\delta}$. These vectors can be categized into two groups depending on the fault produced, with $\boldsymbol{\delta A}, \boldsymbol{\delta B}, \boldsymbol{\delta C}$ corresponding to CESF while $\boldsymbol{2A\delta}, \boldsymbol{2B\delta}$ and $\boldsymbol{2C\delta}$ leading to SESF. It should be noted that $\boldsymbol{2A\delta}, \boldsymbol{2B\delta}$ and $\boldsymbol{2C\delta}$ can be regarded as being equivalent due to their identical three-fold rotational symmetry. Specifically, it can be derived by the following equations considering $\boldsymbol{2AB}$ and $\boldsymbol{2BC}$ are the lattice constants of L1$_2$-Ni$_3$Al:

$$\boldsymbol{2A\delta} = \boldsymbol{2A\delta} - \boldsymbol{2B\delta} + \boldsymbol{2B\delta} = 2(\boldsymbol{A\delta} + \boldsymbol{\delta B}) + \boldsymbol{2B\delta} = \boldsymbol{2AB} + \boldsymbol{2B\delta} = \boldsymbol{2B\delta} \tag{4}$$

Similarly,

$$\boldsymbol{2A\delta} = \boldsymbol{2A\delta} - \boldsymbol{2C\delta} + \boldsymbol{2C\delta} = 2(\boldsymbol{A\delta} + \boldsymbol{\delta C}) + \boldsymbol{2C\delta} = \boldsymbol{2AC} + \boldsymbol{2C\delta} = \boldsymbol{2C\delta} \tag{5}$$

Consequently, there exist four scenarios considering the fault variation ($\Delta S_1 = \boldsymbol{\delta A}, \boldsymbol{\delta B}, \boldsymbol{\delta C}$ or $\boldsymbol{2A\delta}$) in IDL interacting with the super edge dislocation ($\boldsymbol{2BC}$), which have been examined in this study, as illustrated in **Fig. 1(c-e),** distinct from the case the pristine FCC structure where only one scenario exists. [10, 24]

External shear stresses were then applied by imposing equal but opposite forces on to the top and bottom surfaces to drive the partials to interact with the IDL of enclosing

CESF or SESF triggering the planar fault transformation, as shown in Fig.**1(c-e)**. The radii of IDLs were set between 30Å to 105Å to examine the size effect of IDL. The simulations were carried out using the Large-scale Atomic/Molecular Massively Parallel Simulator (LAMMPS) [42]. The interatomic interactions of Ni-Al were described using the embedded atom method (EAM) potential[43] developed by Y. Mishin[44]. This potential fitted to many experimental and first-principles data, and accurately predict the stable ordered intermetallic phases and superlattice stacking fault structures and have been widely used to study the IDLs and super dislocation activities[2, 10, 24]. The energies of various planar faults including SISF, CSF, APB, SESF, and CESF are calculated and listed in **Table I** with the calculation details elaborated in our previous work [2]. The polyhedral template matching method implemented in the OVITO software [45] was employed in our research to analyze the local crystalline structure of IDLs within the $L1_2$ lattice. The relaxation of the system was achieved using the fast inertial relaxation engine (FIRE) methods [46, 47]. This approach has demonstrated remarkable performance and accuracy in identifying local minima under the influence of external stresses.

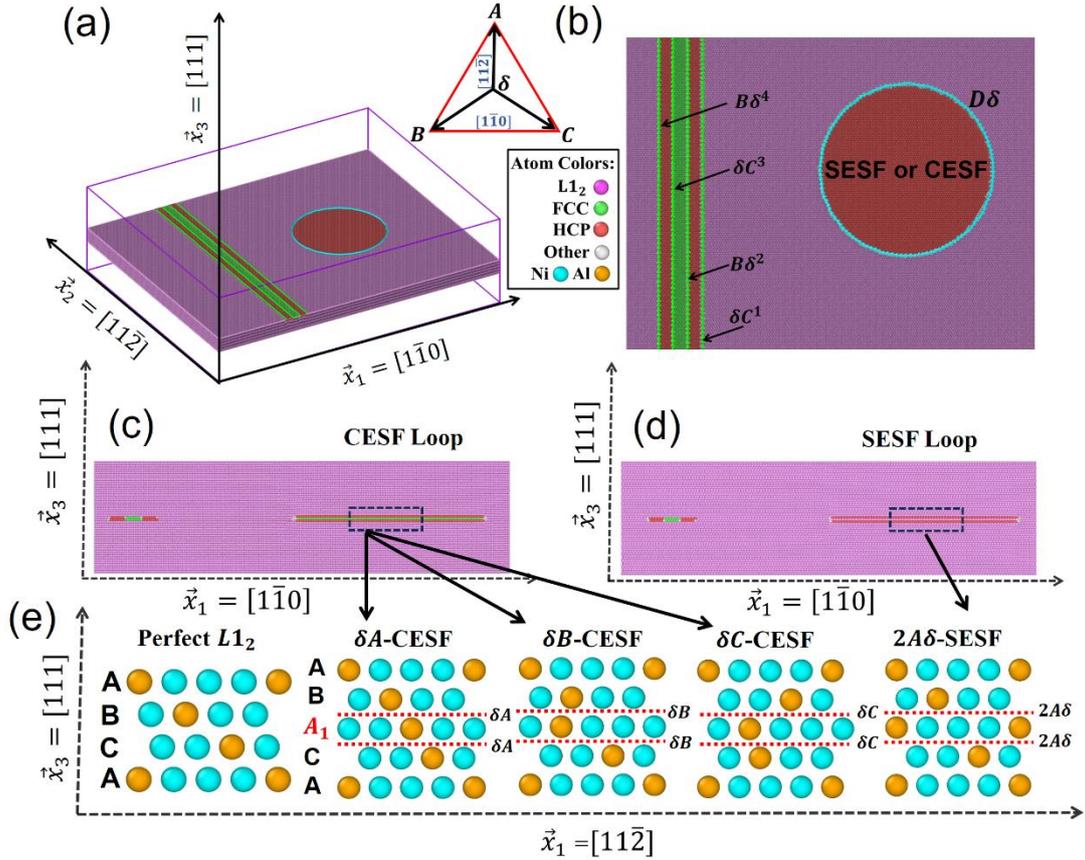

**Fig. 1.** **(a-b)** Atomistic models for simulating a super-edge dislocation ($2BC$), represented by the four constituting partials ($B\delta^4$, $\delta C^3$, $B\delta^2$ and $\delta C^1$), to interact with a single layer IDL of $D\delta$ enclosing a CESF or SESF fault. The corresponding Thompson tetrahedron is shown as an inset figure in **(a)** with the positive dislocation line sense represented by the vector $\vec{x}_2 = [1\bar{1}\bar{2}]$. **(c) and (d)** illustrate the enlarged $(1\bar{1}0)$ sections showing the corresponding stacking sequences of the IDLs enclosing CESF and SESF respectively, while **(e)** shows possible stacking sequences of CESF and SESF with extra atom layer $A_1$ inserted, with the planar fault vectors being $\delta A, \delta B, \delta C$ or $2A\delta$. The dislocation line segments are colored based on their Burgers vectors, with Frank partial, Shockley partial and full dislocation colored cyan, green and bule respectively. Atoms are colored by local lattice environments, identified through the polyhedral template matching method [45].

## 3. Results

### 3.1 Super-edge interacting with IDL of $\delta A$-CESF

As previously illustrated in **Fig. 1**, an IDL may enclose either a CESF (of different fault vectors $\delta A, \delta B$ or $\delta C$) or SESF fault. Starting with IDLs enclosing a CESF with the fault vector being $\delta A$, we examined the interaction process as a super edge was driven towards the IDL, with the results shown in **Fig. 2**. We can see during the initial encounter of the super edge with IDL (see **Fig. 2a-b**), the first partial $\delta C^1$ was blocked at the left side of the IDL, resulting in $(\delta D + \delta C^1)$. However, the blockage of $\delta C^1$ does not seem to inhibit the motion of subsequent partials. In particular, the second partial $B\delta^2$ was observed to continue gliding past the segments of $(\delta D + \delta C^1)$ to get inside the loop and transform the CESF fault into SISF+APB, with the fault-dislocation reaction shown in **Eq. (6)** below:

$$\text{CESF}(\delta A) + B\delta^2 \to \text{CSF}(\delta A) + \text{APB}(BA) \qquad (6)$$

The blockage of $\delta C^1$ at IDL can be attributed to the alternation of the local stacking sequences induced by the inserted atomic layer of IDL, as detailed in **Fig. 3**. According to the Thompson tetrahedron shown in the figure and noting that the gliding occurs on the {111} plane, we see that a fault vector $\delta A$, $\delta B$ or $\delta C$ would transform the original three-fold stacking sequence A-B-C locally to C-A-B, while $A\delta, B\delta$ or $C\delta$ would transform the stacking sequence A-B-C to B-C-A. Thus, the gliding of $\delta A$, $\delta B$ **and** $\delta C$ can produce a complex stacking fault (CSF) in the $L1_2$ lattice (see **Fig. 3b),** but an unstable stacking fault of high energy inside the IDL (see **Fig. 3c**). The high energy barrier of this transformation thus resulted $\delta C^1$ being blocked, while it facilitated the gliding of $A\delta, B\delta$ and $C\delta$ that convert the CESF into the complex of CSF and APB (see **Fig. 3d**). The corresponding energy barrier variation has been illustrated by the planar fault energy transition curves in **Fig. 4**. Similarly, the IDL also blocked $\delta C^3$ but permitted the motion of $B\delta^4$. It could be thus concluded that the slipping sequences of four partials were altered from $B\delta^4$, $\delta C^3$, $B\delta^2$, $\delta C^1$, to $\delta C^3$, $B\delta^4$ $\delta C^1$, $B\delta^2$, when the partials slipped into the region enclosed within the IDL.

With further loading after the entry of $\delta C^1$ and $B\delta^2$ into the IDL, unfaulting occurred through the nucleation and emission of dislocations $A\delta^-\uparrow$ and $A\delta^-$ from the periphery of the loop. Here, $A\delta^-$ would react with $B\delta^2$ and further produce the *D*-Shockley partials of $\delta C^\pm$ (see **Fig. 2c**) to remove the CSF. Here, the superscripts " + " and " − " were introduced to denote Shockley partials sitting below and above the IDL respectively, while the vertical arrow symbol (↑) next to $A\delta^-$ indicates reversal of the dislocation line sense direction [2, 10]. With these notations, the dislocation reactions involved can thus be written in the follows

$$A\delta^- + B\delta^2 \to \delta C^\pm \qquad (7)$$

It should be noted that $A\delta^-\uparrow$ in the left end of IDL will go to the periphery of IDL and react with $\delta C^3$ and the Frank partial dislocation $D\delta\uparrow$ resulting in the dislocation of $BD$ (see **Fig.2d-j**). Here, the D-Shockley partials $\delta C^\pm$ represent $A\delta^-$ and $B\delta^2$ overlapping and being separated by the inserted interstitial layer. This arrangement generates a super jog along the [111] direction, corresponding to $\delta C^\pm$ having a core extended over two adjacent (111) planes [2, 10]. On the other hand, as $\delta C^\pm$ sweeps across the loop to the right end of **IDL,** it would react with the Frank partial dislocation $D\delta$ to produce a dislocation $DC$ (see **Fig. 2d-g**). The associated dislocation reactions are given below:

$$A\delta^-\uparrow + D\delta\uparrow + \delta C^3 \to BD \qquad (8)$$

$$\delta C^\pm + D\delta \to DC \qquad (9)$$

Another thing worth noting is that when the partials were leaving the periphery of IDL, the perfect dislocation $DC$ was replaced by $BD$ to release $B\delta^2$, and at this point the partials were slipping from the IDL region with the stacking sequence permitting the motion of $A\delta, B\delta$ and $C\delta$ to the region of perfect lattice that favors the motion of $\delta A, \delta B$ or $\delta C$. As a result, the slipping sequences of four partials were altered again from $\delta C^3, B\delta^4 \delta C^1, B\delta^2$ to $B\delta^4, \delta C^3, B\delta^2, \delta C^1$, recovering the original sequence (c.f., **Fig. 2a**). Additionally, we noted that during the process only $B\delta^4$, $B\delta^2$ and $\delta C^1$ sheared the IDL, while the looping mechanism occurred for $\delta C^3$ that contributed to unfaulting the IDL to a full dislocation $BD$, as indicated in **Eq. (8)**

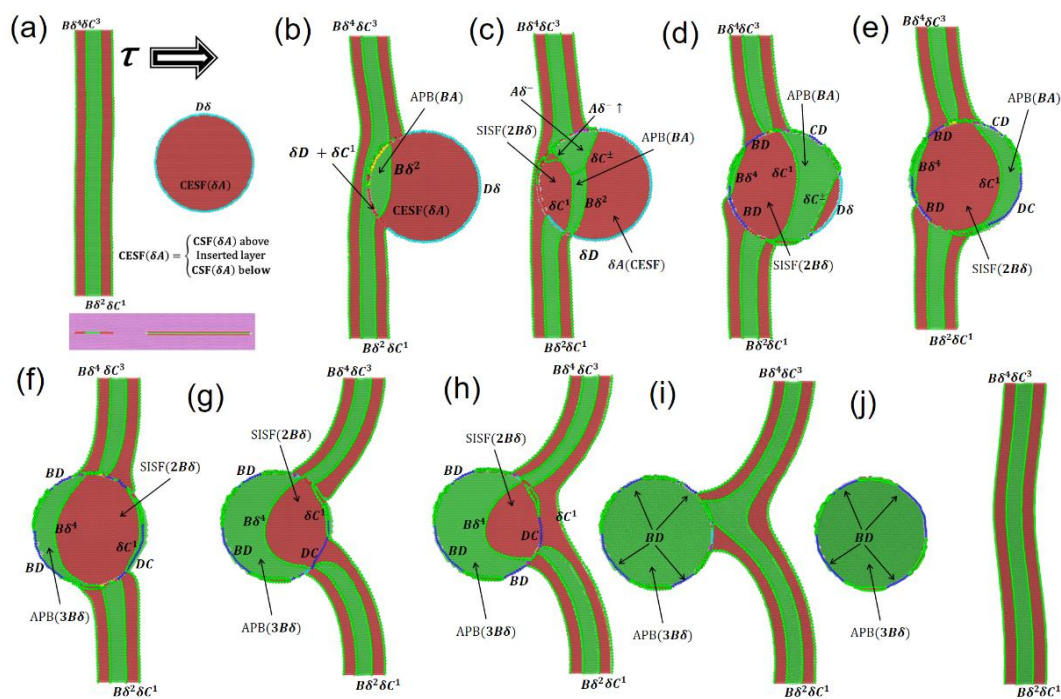

**Fig. 2. (a)** The initial atomistic model where the super-edge dislocation ($2BC$), in its dissociated form, were driven to interact with an IDL of the $\delta A$-CESF type and a radius being 9nm. **(b)** $\delta C^1$ was blocked at the left side of the IDL **(c)** after the entry of $\delta C^1$ and $B\delta^2$ into the IDL, unfaulting occurred from the periphery of the loop. **(d)** $A\delta^-$ would with and $B\delta^2$ and further produce the $D$-Shockley partials of $\delta C^\pm$. **(e-f)** the dislocation reactions of $A\delta^- \uparrow + D\delta \uparrow + \delta C^3 \rightarrow BD$ and $\delta C^\pm + D\delta \rightarrow DC$ **(g-h)** the perfect dislocation $DC$ was replaced by $BD$ to release $B\delta^2$. **(i-j)** the slipping sequences of the four partials, originally of $B\delta^4, \delta C^3, B\delta^2, \delta C^1$, change into $\delta C^3, B\delta^4 \delta C^1, B\delta^2$ when they sheared into the IDL, and then revert back to the original as they slip out of the IDL.

**Table I:** The energies of various planar faults in L1$_2$-Ni$_3$Al, as predicted by the EAM potential developed by Y. Mishin [44]

| Planar fault energy (mJ/m$^2$) | | | | |
|---|---|---|---|---|
| $\gamma_{SISF}$ | $\gamma_{APB}$ | $\gamma_{CSF}$ | $\gamma_{SESF}$ | $\gamma_{CESF}$ |
| 50 | 250 | 200 | 54 | 427 |
| Change in fault energy (mJ/m$^2$) | | | | |
| CESF → CSF + APB | | CESF → CSF | SESF → APB + SISF | |
| 35 | | -232 | 202 | |

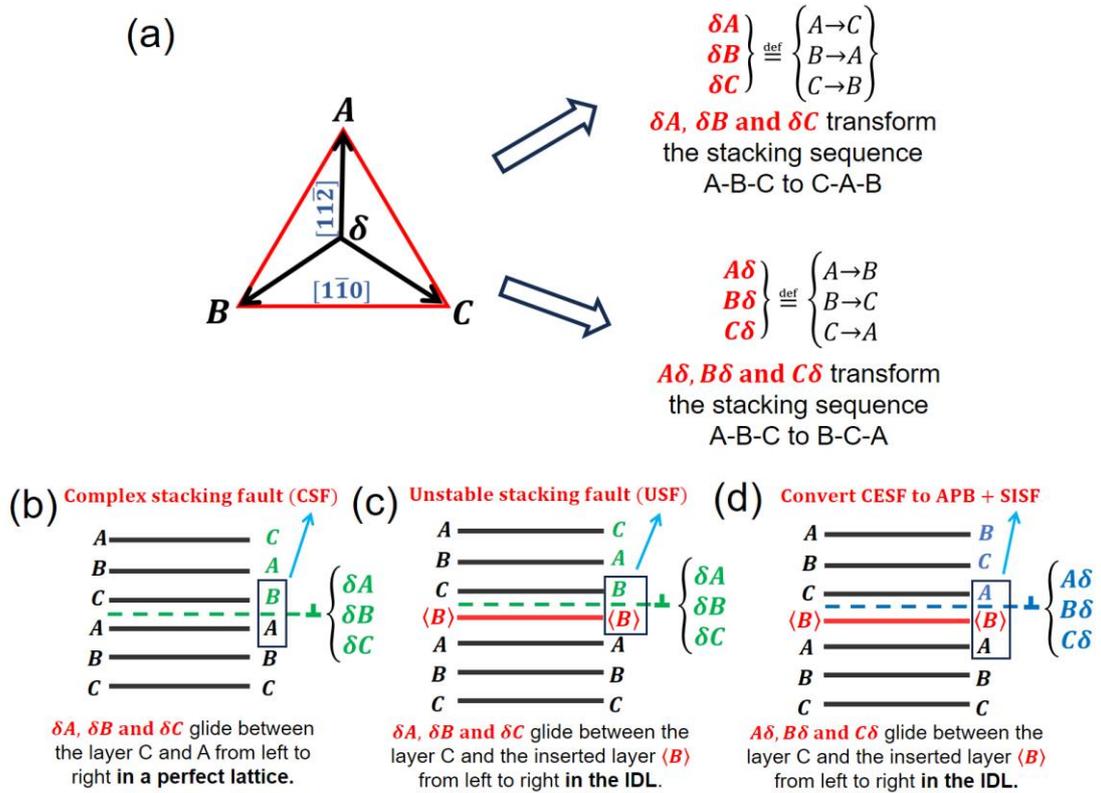

**Fig.3. (a)** The Thompson tetrahedron opened up at corner D with the corresponding Shockley partials ($\delta A, \delta B, \delta C$) illustrated, also showing the transformation of the stacking sequence by different fault vector. **(b)** shows the gliding of one of $\delta A, \delta B$ and $\delta C$ results in the formation CSF in a perfect lattice. **(c)** demonstrates the gliding of one of $\delta A, \delta B$ and $\delta C$ leads to the formation of unstable stacking fault in **CESF** in the IDL. **(d)** reveals that the gliding of one of $A\delta, B\delta$ and $C\delta$ facilitates the transformation of CESF to APB and SISF in the IDL.

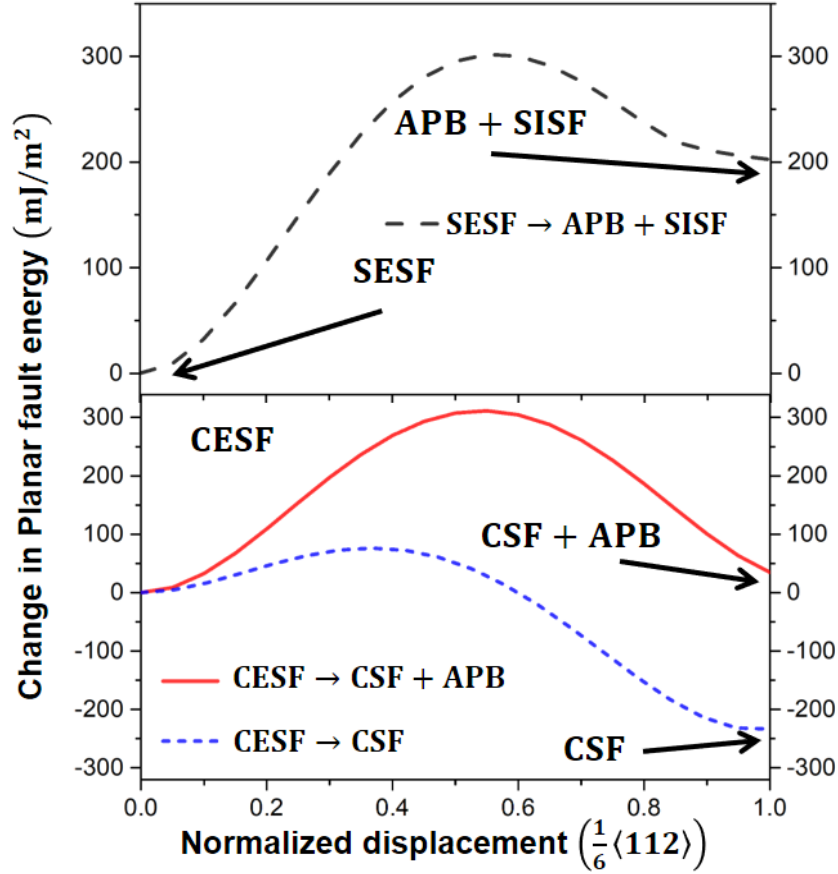

**Fig.4.** The planar fault energy variation curves, along the slipping direction of $B\delta^2$, corresponding to the transformation routes of $\text{CESF}(\delta A) + B\delta^2 \rightarrow \text{CSF}(\delta A) + \text{APB}(BA)$ (solid red), $\text{CESF}(\delta B) + B\delta^2 \rightarrow \text{CSF}(\delta B)$ (dashed blue) and $\text{SESF}(2A\delta) + B\delta^2 \rightarrow \text{APB}(3B\delta) + \text{SISF}(2A\delta)$ (dashed black).

### 3.2 Super-edge interacting with IDLs of $\delta B$-CESF and $\delta C$-CESF

In addition to the case IDL CESF loop shown in **Fig. 3**, the interaction between a super-edge with the IDLs enclosing the $\delta B$-CESF and $\delta C$-CESF faults were also examined. **Fig. 5** shows the interaction mechanism between the super-edge dislocations and the CESF loop having the fault vector of $\delta B$. From the figure, we can see that the slipping sequences of four partials has also been converted from $B\delta^4, \delta C^3, B\delta^2, \delta C^1$ to $\delta C^3, B\delta^4 \, \delta C^1, B\delta^2,$ when these partials slipped into the region enclosed within the IDL, similar to that previously shown in **Fig. 2**. As previously explained, such conversion in the slipping sequences of partials is due to the slipping of $\delta A$, $\delta B$, or $\delta C$ slipping leading to unstable high-energy faults.

However, despite some similarity, clear distinction can be observed in the reaction process as we compare **Fig. 5** with **Fig. 2**. In the case shown in **Fig. 5**, the unfaulting reaction occurred through the nucleation of $\boldsymbol{B\delta^-}\uparrow$ and $\boldsymbol{B\delta^-}$ in the IDL, instead of $\boldsymbol{A\delta^-}\uparrow$ and $\boldsymbol{A\delta^-}$ as previously shown in **Fig. 2**. This change occurred because the Burgers vectors of the nucleated dislocations needed to counteract the displacement vectors of planar faults within the IDL. Specifically, in **Fig. 5**, we see that in the early stage, the two partials $\boldsymbol{B\delta^-}$ and $\boldsymbol{B\delta^2}$ removed the CESF with two-layer CSFs (see **Fig. 5a-d**), resulting in the formation of $\boldsymbol{AD}$, which was then followed by the gliding of $\boldsymbol{\delta C^1}$ to produce another CSF (see **Fig. 5e-f**). The corresponding energy barrier of the transition of CESF → CSF was given in **Fig. 4**. It should be noted that part of $\boldsymbol{B\delta^4}$ got annihilated by $\boldsymbol{B\delta^-}\uparrow$, when it contacted into the IDL reacting with $\boldsymbol{AD}$ through $\boldsymbol{AD} + \boldsymbol{B\delta^-}\uparrow \rightarrow \boldsymbol{D\delta}\uparrow + \boldsymbol{\delta C^1}$ (see **Fig. 5c-d**). The produced $\boldsymbol{\delta C^1}$ then swept across the IDL to generate the CSF again. However, since $\boldsymbol{\delta C^3}$ cannot shear into the IDL due to the high energy barrier of producing unstable stacking fault (see **Fig. 3**), as previously explained, consequently, further shearing then drove the looping mechanism to occur for $\boldsymbol{\delta C^3}$ to generate the $\boldsymbol{\delta C^3}+\boldsymbol{\delta D}$ (see **Fig. 5g-h**). The key dislocation interactions involved in the process shown in **Fig. 5** are the follows:

$$\boldsymbol{B\delta^2} + \text{CESF}(\boldsymbol{\delta B}) \rightarrow \text{CSF}(\boldsymbol{\delta B}) \tag{10}$$

$$\boldsymbol{B\delta^-} + \text{CSF}(\boldsymbol{\delta B}) \rightarrow 0 \tag{11}$$

$$\boldsymbol{B\delta^-}\uparrow + \boldsymbol{D\delta}\uparrow + \boldsymbol{\delta C^1} \rightarrow \boldsymbol{AD} \tag{12}$$

$$\boldsymbol{B\delta^4} + \boldsymbol{B\delta^-}\uparrow \rightarrow 0 \tag{13}$$

$$\boldsymbol{\delta C^3} + \boldsymbol{\delta D} \rightarrow (\boldsymbol{\delta C^3} + \boldsymbol{\delta D}) \tag{14}$$

The above results demonstrate that $\boldsymbol{B\delta^4}$, $\boldsymbol{B\delta^2}$ and $\boldsymbol{\delta C^1}$ sheared the IDL, while $\boldsymbol{\delta C^3}$ got past the IDL by the looping mechanism. The segments of $\boldsymbol{B\delta^4}$ in the IDL was annihilated with the $\boldsymbol{B\delta^-}\uparrow$ nucleated during shearing the IDL.

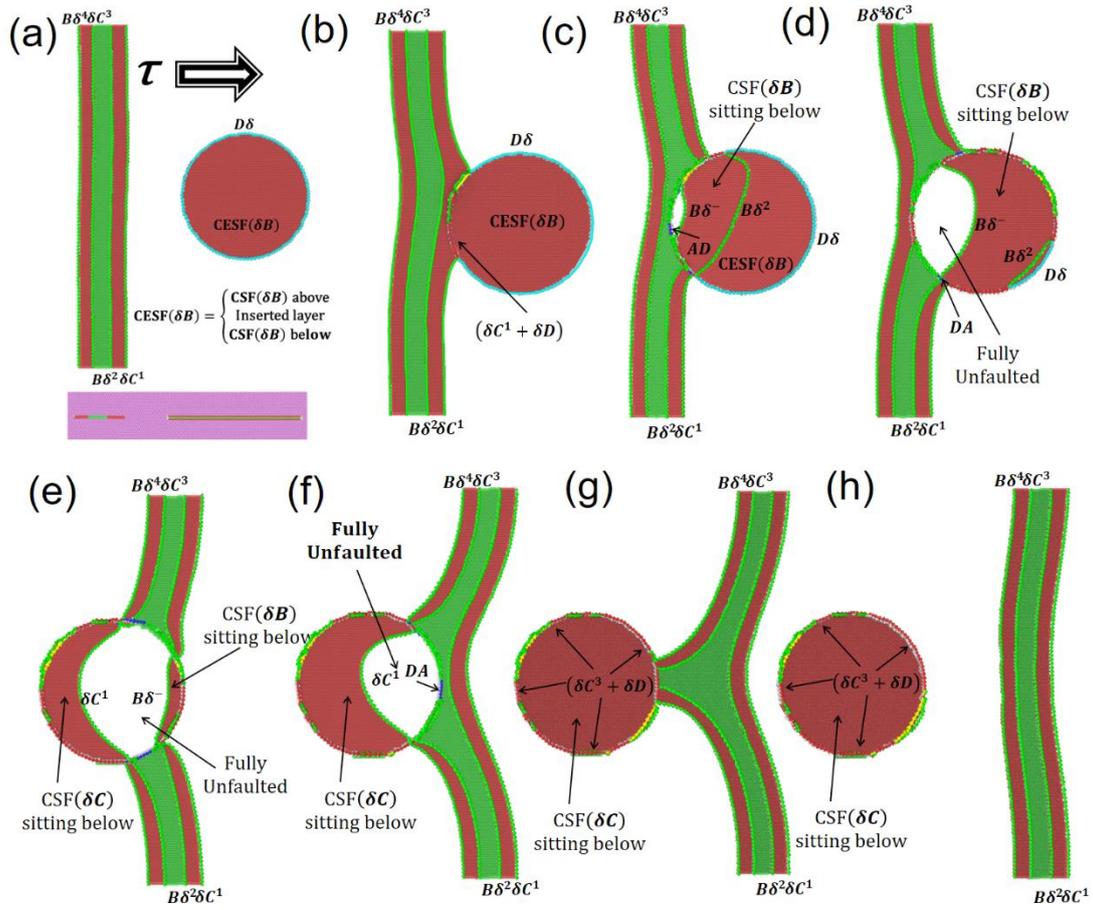

**Fig.5. (a)** The initial atomistic model where four moving Shockley partials ($B\delta^4$, $\delta C^3$, $B\delta^2$ and $\delta C^1$) dissociated from a super-edge dislocation ($2BC$) were driven to interact with an IDL of $\delta B$-CESF type having a radius being 9nm. **(b)** $\delta C^1$ was blocked at the left side of the IDL **(c-d)** The unfaulting reaction occurred through the nucleation of $B\delta^- \uparrow$ and $B\delta^-$ in the IDL with the reaction of $B\delta^- \uparrow + D\delta \uparrow + \delta C^1 \to AD$. **(e)** $B\delta^4$ got annihilated with $B\delta^- \uparrow$, when it contacted into the IDL reacting with **AD** by $AD + B\delta^- \uparrow \to D\delta \uparrow + \delta C^1$. **(f)** $\delta C^1$ swept across the IDL to generate the CSF. **(g-h)** $B\delta^4$, $B\delta^2$ and $\delta C^1$ sheared the IDL, while $\delta C^3$ moved forward leaving the Orowan loop around the IDL.

**Fig.6**. presents the other case of the super-edge dislocation interacting with the CESF loop having the fault vector of $\delta C$. Again, we can see that the $\delta C^3$ partial was blocked but the other partials, $B\delta^4$, $\delta C^1$ and $B\delta^2$ can be driven to shear the IDL successively. However, the detailed reaction process exhibits some difference from what's shown in **Fig. 2** and **Fig. 5**. The unfaulting of the top CSF of the CESF in the IDL mandate synergetic actions of $B\delta^4$, $\delta C^1$ and $B\delta^2$ through the following reaction (see **Fig. 6a-e**). It yields the fault vector 2**BC** which equals the lattice translation vector necessary to restore lattice perfection. This differs significantly from dislocation reactions identified in **Figs. 2 and 5**, where only a single partial dislocation is required for unfaulting the upper CSF.

$$B\delta^4 + \delta C^1 + B\delta^2 + \text{CSF}(\delta C) \to 2\mathbf{BC} \quad (15)$$

The other CSF of the CESF was unfaulted by the nucleation and slipping of $\delta C^- \uparrow$ and $\delta C^-$ inside the IDL, where $\delta C^- \uparrow$ reacted with $D\delta$ and $B\delta^4$ resulting in the formation of $DA$, as illustrated in the below:

$$\delta C^- \uparrow + B\delta^4 + D\delta \to DA \quad (16)$$

But we noted that $\delta C^-$ and $\delta C^- \uparrow$ would disappear finally under the effects of external shear, and it is interesting to see that $\delta C^3$ again looped the IDL to generate the $\delta C^3 + \delta D$ due to the high energy barrier to trigger the unstable stacking fault showing great similarity to the cases of **Fig. 2** and **Fig. 4.**

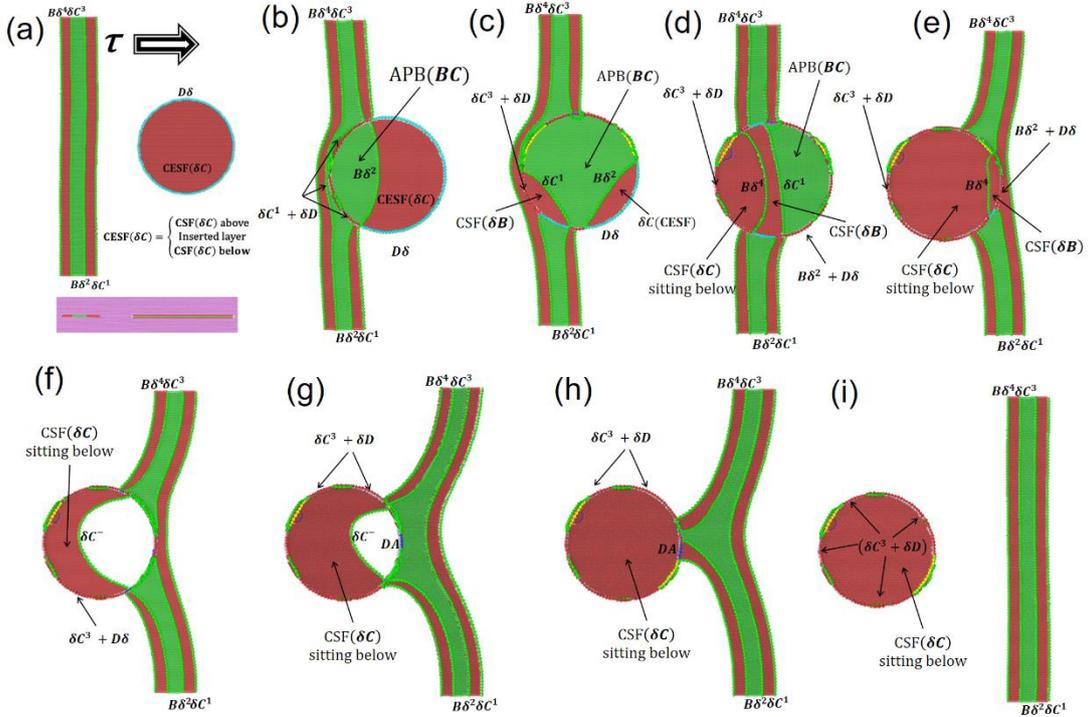

**Fig.6. (a)** The initial atomistic model where four moving Shockley partials ($B\delta^4$, $\delta C^3$, $B\delta^2$ and $\delta C^1$) dissociated from a super-edge dislocation (2**BC**) were driven to interact with an IDL of $\delta C$-CESF type having a radius being 9nm. **(b-c)** The first partial $\delta C^1$ was initially blocked on the left side of the IDL then slipped into the IDL following the slip of $B\delta^2$. **(e-f)** The third partial $\delta C^3$ was blocked on the left side of the IDL but $B\delta^4$ moved into the IDL. **(g-h)** The nucleation and slipping of $\delta C^- \uparrow$ and $\delta C^-$ inside the IDL unfaulted the CSF of the CESF. **(i)** $B\delta^4$, $B\delta^2$ and $\delta C^1$ sheared the IDL, while $\delta C^3$ moved forward leaving the Orowan loop around the IDL.

### 3.3 Super-edge interacting with IDL of $2A\delta$-SESF

Now we turn our attention to the IDLs enclosing SESF, produced by the fault vectors $2A\delta, 2B\delta$ and $2C\delta$. As previously mentioned, due to the three-fold rotational symmetry, these three (fault vector) cases are equivalent, and thus we only considered the IDL of the fault vector $2A\delta$. It is important to note that the fault vectors $2A\delta, 2B\delta$ and $2C\delta$ lead to the formation of SESF with a formation energy of 54 mJ/m², much lower than that of the CESF loop, of a formation energy of 427 mJ/m² (see **Table I).** Thus, the IDLs enclosing SESF are much more stable structures, and it is expected that no unfaulting would be triggered for them.

The interaction process between the super-edge dislocation and the SESF loop of a planar fault vector of $2\delta A$ is illustrated in the **Fig. 7.** Following the same rules, the slipping sequences of four partials has been changed in the IDL region but restored after leaving the IDL. However, differences have been noted that $\delta C^3, B\delta^4, \delta C^1$ and $B\delta^2$ all sheared the IDL successively with different energy barriers and resistance, and no looping mechanism was observed. The planar fault has been first changed from SESF($2A\delta$) to APB($3B\delta$) + SISF($2A\delta$) when $B\delta^2$ slip across the IDL. It needs to overcome an energy barrier of 202 mJ/m². As evidenced in Fig. 4, this initial higher energy barriers during the planar fault transformation from SESF to SISF+APB prevented the nucleation of Shockley dislocations to induce the unfaulting of SESF loop, in contrasts with the observed scenarios of CESF loops. Subsequently, the motion of $\delta C^1$ transformed it to CSF($\delta A$) + SISF($2A\delta$), which encountered a negative energy barrier of -48 J/m². This negative value indicates a thermodynamically spontaneous entry. The followed $B\delta^4$ further altered this planar fault to APB($BA$) + SISF($2A\delta$) with an energy barrier of 48 mJ/m², which is relatively lower than the initial 202 mJ/m² barrier for the formation of second APB**.** Finally, $\delta C^3$ sheared into the IDL, and then revert back to the original as it slipped out of the IDL with a negative energy barrier of -202 mJ/m² indicating a highly spontaneous process.

The planar fault transformations triggered by the four partials could be concluded in the following.

$$B\delta^2 + \text{SESF}(2A\delta) \rightarrow \text{APB}(3B\delta) + \text{SISF}(2A\delta) \qquad (17)$$

$$\delta C^1 + \text{APB}(3B\delta) + \text{SISF}(2A\delta) \rightarrow \text{CSF}(\delta A) + \text{SISF}(2A\delta) \qquad (18)$$

$$B\delta^4 + \text{CSF}(\delta A) + \text{SISF}(2A\delta) \rightarrow \text{APB}(BA) + \text{SISF}(2A\delta) \qquad (19)$$

$$\delta C^3 + \text{APB}(BA) + \text{SISF}(2A\delta) \rightarrow \text{SESF}(2A\delta) \qquad (20)$$

The sequence of transformations shows that the IDLs of CESF and SESF both undergo complex evolution pathways during interplaying with super edge dislocation, involving different types of planar faults. Each transformation has its own energetic considerations, affected by the faulted vectors after the Shockley partials sweeping the IDL.

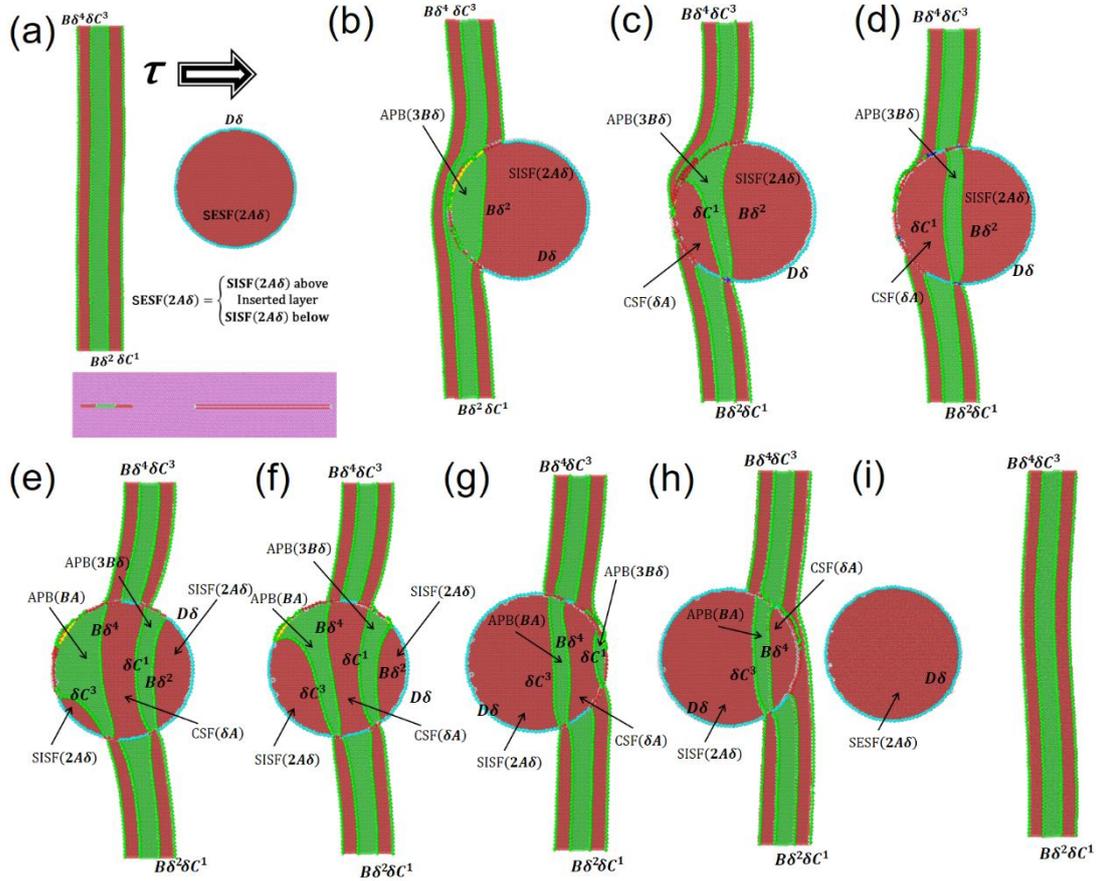

**Fig.7. (a-i)** shows the evolution processes as four moving Shockley partials ($B\delta^4$, $\delta C^3$, $B\delta^2$ and $\delta C^1$) dissociated from a super-edge dislocation ($2BC$) were driven to interact with an IDL of **2A$\delta$-SESF** type with a radius being 9nm. **(a-d)** $\delta C^1$ and $B\delta^2$ slip into the IDL to produce APB($3B\delta$) and CSF($\delta A$). **(e-f)** $\delta C^3$ and $B\delta^4$ moved into the IDL inducing the fault of APB($BA$) and revert the fault back to the original SISF($2A\delta$) **(g-i)** $\delta C^3$, $B\delta^4$, $\delta C^1$ and $B\delta^2$ all sheared the IDL successively restoring the IDL fault to SESF($2A\delta$).

## 4. Discussions

### 4.1 Size effect of reaction mechanism and structure evolution

In our results presented above, for simplicity and consistency, the IDLs were set to have a fixed radius of 9nm. Despite this previous simulation set-up producing good results to elucidate the diverse interaction mechanisms between the super edge dislocation and IDLs, it does not account for the potential effect from the variation in the IDL radius.

As demonstrated in the previous studies [25, 48], such variation can yield a size effect influencing the critical stress and obstacle strength. In this regard, we carried out additional simulations where the radii of IDLs were varied from 30Å to 105Å. Interestingly, we found that the size effect critically depends on the fault enclosed within the IDL, as elaborated below.

For the cases of IDLs of **$\delta B$**-CESF **and 2$A\delta$**-SESF, it was discovered that the reaction mechanism and structure evolution during the dislocation-planar fault interaction remained the same, independent of the IDL radius (**see Section 1-2 in supporting information**). However, for the IDLs of **$\delta A$**-CESF **and $\delta C$**-CESF, a clear size effect was observed. Starting with the IDL of **$\delta A$**-CESF, it was found that the structure evolution events would be rendered different if the IDL radius drops below a threshold value (~ 3.5nm). This is illustrated in **Fig.8**, where the IDL assumes a small radius of 3nm. Comparing with the process previously illustrated in **Fig.2** for the IDL with a radius of 9nm, we note that the initial events remain similar, i.e., the motion of dislocation partials from matrix to IDL resulting in the full dislocation **BD** and **DC.** However, unlike what's shown in **Fig. 2,** the two full dislocations would dissociate along certain orientations, specifically those where the adjacent {111} and planes intersect the original (111) plane and formed extended CSF and APB, as illustrated in **Fig. 8b**. Consequently, under the external shear stress, **BD** and **DC** and their dissociated partials would continually cross slip and combine into **BC** resulting in the formation of pyramid structure in **Fig. 8c**. It should be noted that although this pyramid structure has similar shape to the stacking fault tetrahedron (SFT) in pristine FCC metals, their formation mechanisms differ significantly. The origin of SFT is widely attributed to the Silcox-Hirsch mechanism, in which a Frank partial dissociates into a low-energy stair-rod dislocation and a Shockley partial on intersecting slip planes among three {111} planes of the Thompson tetrahedron[49, 50]. These partials then attract each other in pairs, forming another set of stair-rods that result in the SFT. While for the pyramid structure in **Fig. 8c**, the reaction of two full dislocations is responsible for its formation. This is a good example reflecting the substantial differences in irradiation-induced microstructure evolution between materials systems of pristine FCC and $L1_2$ lattices.

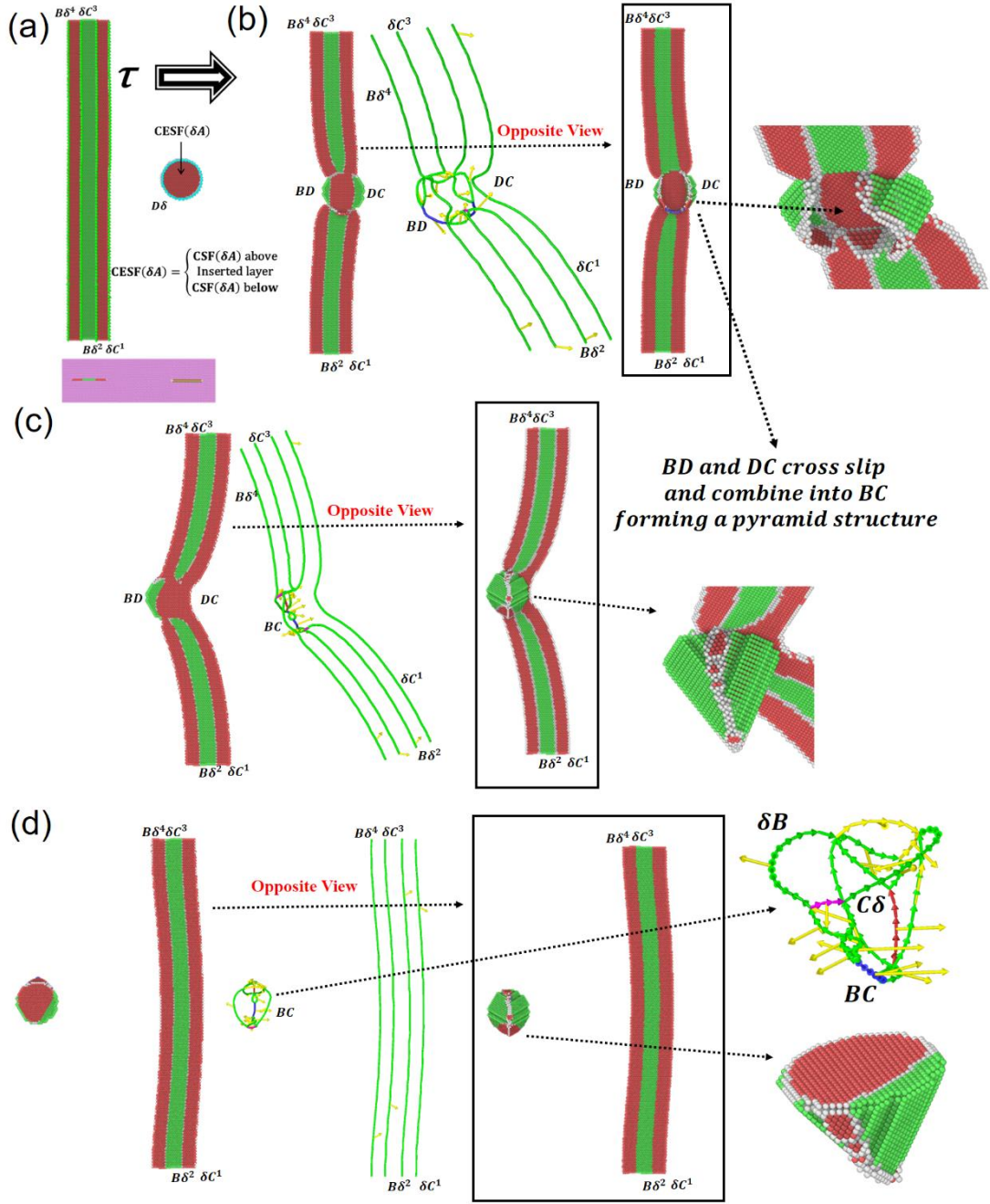

**Fig. 8. (a-b)** show the evolution processes of four moving Shockley partials ($B\delta^4$, $\delta C^3$, $B\delta^2$ and $\delta C^1$) interacting with an interstitial dislocation loop of $\delta A$-CESF type with a smaller radius being 3nm. **(a-b)** $B\delta^4$, $\delta C^3$, $B\delta^2$ and $\delta C^1$ slipped into the IDL and produced two full dislocations $BD$ and $DC$. **(c)** The small loop facilitates the formation of pyramid structure resulting from the dissociation reactions of $BD$ and $DC$ along certain orientations combining into $BC$, specifically those where the adjacent {111} planes intersect the original [111] plane. **(d)** Four partials moved out of the IDL and left the pyramid structure with the ledges being the dislocation line of $BC$ or its dissociated partials. For clarity, the identified dislocation lines are separated from the atomic configurations with the corresponding Burgers vectors indicated. The opposite views are also given in **(b-d)**.

Additionally, for the scenario of $\boldsymbol{\delta C}$-CESF type loop, **Fig. 9** presents the super edge dislocation interacting with the IDL of radius being 3nm. We can see that the interactions (**Fig. 9a-e**) show great similarity to the case of radius being 9nm (**Fig. 6a-f**). As noted in **Fig. 9**, $\boldsymbol{\delta C^1}$ was initially blocked on the left side of the IDL but $\boldsymbol{B\delta^2}$ slipped into the IDL to produce APB($\boldsymbol{BC}$). Subsequently, after pushing $\boldsymbol{\delta C^1}$ out, $\boldsymbol{\delta C^3}$ was blocked on the left side of the IDL. Meanwhile, $B\delta^4$ moved into the IDL unfaulting the above CSF to be perfect. Finally, $\boldsymbol{B\delta^4}$, $\boldsymbol{B\delta^2}$ and $\boldsymbol{\delta C^1}$ sheared out of the IDL. $\boldsymbol{\delta C^3}$ moved forward leaving the Orowan loop around the IDL. Despite that the process shown in **Fig. 9** sharing similarity to that in **Fig. 6**, there is also distinction with respect to the nucleation and slipping of $\delta C^- \uparrow$ and $\delta C^-$, which are not energetically favorable for IDLs of small radii. This is because the reduction of planar faults contained in the smaller CESF loop cannot compensate for the increase in strain energy stored in the growing Shockley loop. As a result, the complete unfaulting of $\boldsymbol{\delta C}$-CESF loop becomes impossible. This simplifies the interaction between incoming dislocation lines and the $\boldsymbol{\delta C}$-CESF loop, as depicted in **Fig. 9**, which highlights the critical importance of the energy barrier associated with superlattice planar faults in dictating the structural evolution of dislocation loop/line complex. Additionally, compared with the case of $\boldsymbol{\delta A}$-CESF type with a 3nm radius, it was observed that the dissociation reactions of $(\boldsymbol{\delta C^1} + \boldsymbol{\delta D})$, $(\boldsymbol{\delta C^3} + \boldsymbol{\delta D})$, $(\boldsymbol{B\delta^2} + \boldsymbol{\delta D})$ and $(\boldsymbol{B\delta^4} + \boldsymbol{\delta D})$ along the dislocation loop would not lead to the formation of pyramid structure. This suggests that both the geometrical configurations and types of superlattice planar faults play a critical role in determining the resultant structural formations from dissociation reactions.

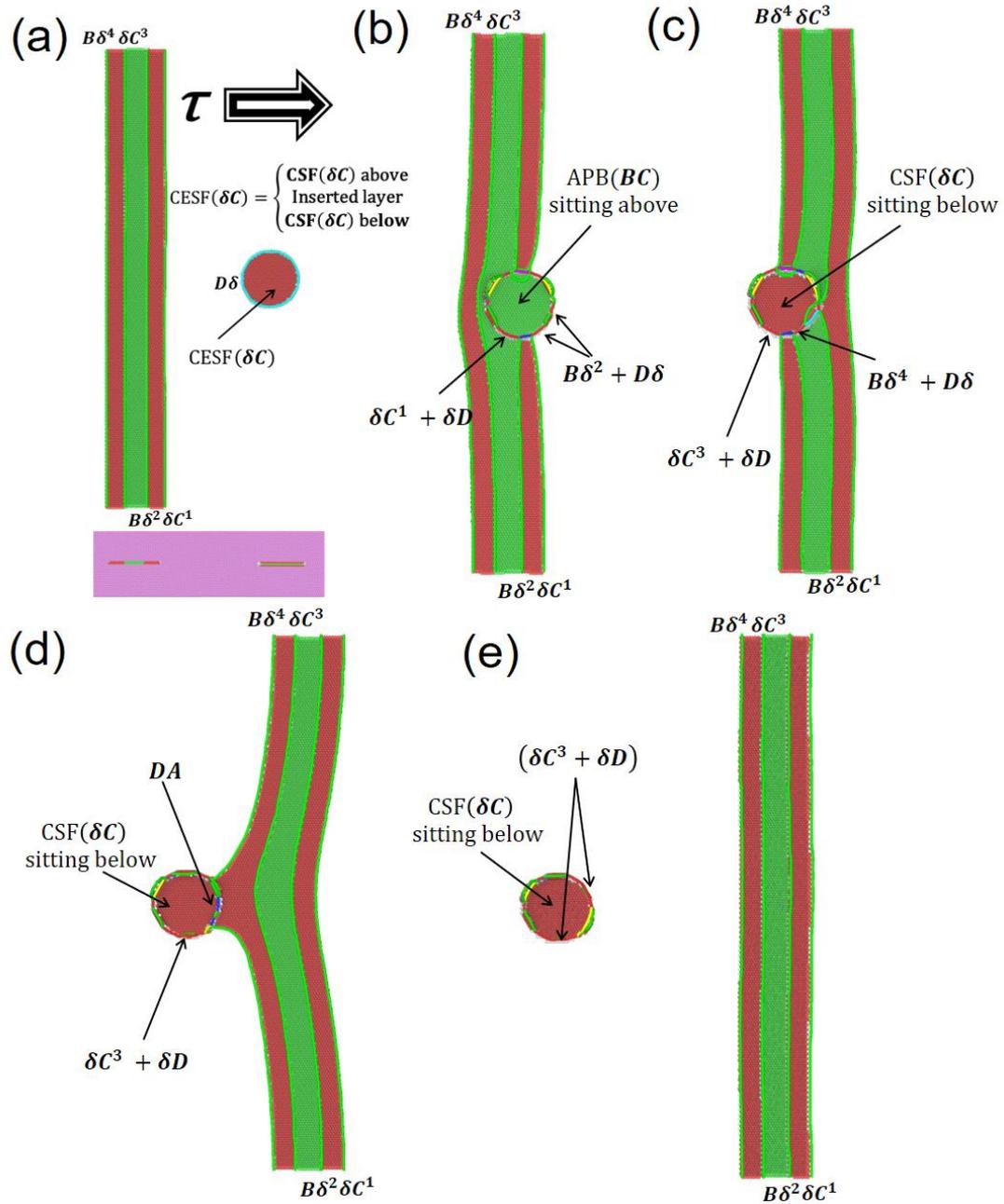

**Fig.9. (a-e)** shows the evolution processes as four moving Shockley partials ($B\delta^4$, $\delta C^3$, $B\delta^2$ and $\delta C^1$) were driven to interact with an IDL of $\delta C$-CESF type having a radius being 3nm. CESF($\delta C$) has two CSFs($\delta C$) sitting below and above the inserted atom layer. **(a-b)** $\delta C^1$ was initially blocked on the left side of the IDL but $B\delta^2$ slipped into the IDL to produce APB($BC$). **(c)** After pushing $\delta C^1$ out, $\delta C^3$ was blocked on the left side of the IDL. Meanwhile, $B\delta^4$ moved into the IDL unfaulting the above CSF to be perfect. **(d)** $B\delta^4$, $B\delta^2$ and $\delta C^1$ sheared out of the IDL. **(e)** $\delta C^3$ moved forward leaving the Orowan loop around the IDL.

## 4.2 Size effect of CRSS

Besides the size effect on the reaction mechanism and structure evolution, the IDL size also has implication to the resistance to the motion of the super-edge. The critical shear stress, denoted as $\tau_{cs}$, required for the super-edge dislocation to move past the IDL of four types planar faults, either via shearing or looping mechanism have been further determined as a function of the IDL radius, shown in **Fig. 10.** We see that for all cases $\tau_{cs}$ monotonically increases as the radii of IDLs increase. Such variation in CRSS is expected from the classical dislocation theory (i.e., Peach–Kohler force) [51], as a larger radius effectively indicating a bigger obstacle size, thus larger resistance. However, it can be noted that for the same IDL radius, the $\tau_{cs}$ value show considerable variation for IDLs of different superlattice planar faults. Ranking the critical stresses for the different types of IDLs, we see that generally $\tau_{cs}$(CESF- $\boldsymbol{\delta A}$) > $\tau_{cs}$(CESF- $\boldsymbol{\delta B}$) > $\tau_{cs}$(CESF- $\boldsymbol{\delta C}$) > $\tau_{cs}$(SESF-$\boldsymbol{2A\delta}$). In particular, we note that the IDLs of CESF-$\boldsymbol{\delta A}$ and SESF types exhibiting the highest and lowest resistance to the motion of the super-edge dislocation, respectively. This can be attributed to the fact that the super-edge dislocation triggered the nucleation of partial dislocations and the subsequent local unfaulting processes of CESF, whereas no unfaulting occurred for SESF due to the high energy barrier of planar fault transformation.

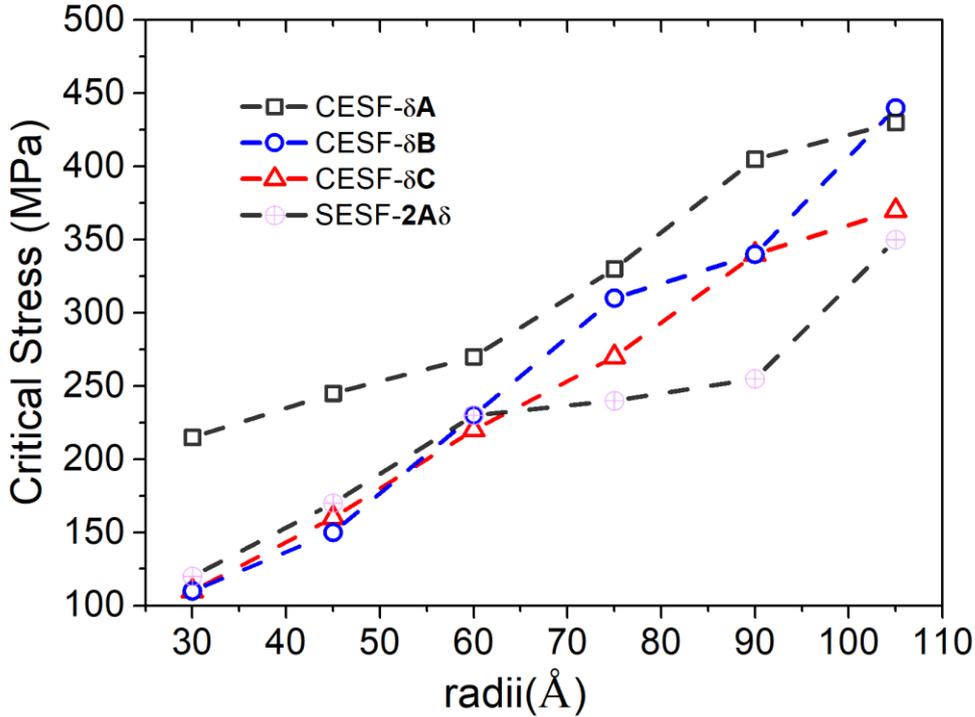

**Fig.10.** The critical resolved shear stress (CRSS) required for the super-edge dislocation to move past **the IDL** enclosing four types planar faults: **CESF- $\boldsymbol{\delta A}$, CESF- $\boldsymbol{\delta B}$, CESF- $\boldsymbol{\delta C}$ and SESF-$\boldsymbol{2A\delta}$.**

## 5. Conclusion

In summary, the present study presented a comprehensive investigation of the interactions between the super-edge dislocation of Burgers vector $\langle 110 \rangle$ and interstitial dislocation loops (IDLs) in irradiated $L1_2$-$Ni_3Al$ through MD simulations. Accounting for the symmetry breakage in the $L1_2$ lattice, four types of IDLs were considered, enclosing either a complex extrinsic stacking fault (CESF) or a superlattice extrinsic stacking fault (SESF), of different fault vectors. The key findings are listed in the follows.

1. The super-edge dislocation interacts with an IDL in terms of its four constituting Shockley partials. The detailed dislocation reactions and structure evolution events as the four partials interact with different IDLs were fully revealed.

2. The slip sequence of the partials constituting the super-edge is found to change as they slip through an IDL, but revert back after they slip past the IDL. This is attributed to the alternation of the local stacking sequences induced by the inserted atomic layer of IDL. Consequently, the slip of the partials within the IDL leads to different transformations of superlattice planar faults and associated energy barriers, which dictates the change of slip sequence of the partials.

3. As the super-edge slips through IDLs enclosing a CESF fault, three of its four partials shear through but one loops around the IDL, where significant energy barrier makes all partials shearing through IDL energetically unfavorable. Partial unfaulting processes were observed for the resulting IDLs with sessile configurations that transform the CESF containing two CSFs to CSF with a lower energy state.

4. On the other hand, as the super-edge slips through an IDL enclosing the SESF fault, all of its partials shear through the IDL successively, and no looping mechanism was observed. This highlighted the dynamic and complex nature of IDLs in CESF and SESF structures when subjected to the shearing of a super-edge dislocation. The resulting transformations were driven by specific energetic factors related to the resulting fault vectors induced by partial dislocations.

5. When the radius of IDLs drops below a certain threshold, a pronounced size effect was observed in their interactions with super-edge dislocations. This phenomenon was attributed to the dissociation of full dislocations along particular orientations, where adjacent {111} planes intersect the original (111) plane, leading to the formation of CSFs and APBs. The resulting faults can produce pyramid-like structures, depending on the geometric configurations and the types of the superlattice planar faults formed by dissociation reactions.

6. The critical shear stress required for the super-edge dislocation to move past the IDL is found to increases monotonically as the IDL size increases. Further, the critical stress is dependent on both the type of fault enclosed in the IDL and the fault vector.

Our findings thoroughly elucidated the atomic details underlying the interaction between a super edge dislocation and IDLs in the $L1_2$ $Ni_3Al$ system, demonstrating the necessity to consider breakage of the 3-fold rotational symmetry of planar faults and their transformations triggered by different Shockley partials, in sharp contrast to the pristine FCC system. With IDLs being an important category of irradiation-induced defects, the present study provides important mechanistic insights contributing to accurate understanding of plastic deformation and irradiation hardening in $L1_2$-$Ni_3Al$, as well as other $L1_2$ systems.

## Acknowledgement

We greatly thank the financial support from National Natural Science Foundation of China (NSFC Grant No. 12002277), Fundamental Research Funds for the Central Universities (Grant No. 06010/G2020KY05111), and Natural Sciences and Engineering Research Council of Canada (NSERC) Discovery grant (RGPIN-2023-03628).